# High Performance Molybdenum Disulfide Amorphous Silicon Heterojunction Photodetector

Mohammad R. Esmaeili-Rad and Sayeef Salahuddin*

Department of Electrical Engineering and Computer Sciences, University of California, Berkeley, California 94720, United States

Emails: {esmaeili , sayeef} @eecs.berkeley.edu

One important use of layered semiconductors such as molybdenum disulfide ($MoS_2$) could be in making novel heterojunction devices leading to functionalities unachievable using conventional semiconductors. Here we demonstrate an ultrafast metal-semiconductor-metal heterojunction photodetector, made of $MoS_2$ and amorphous silicon (a-Si), with rise and fall times of about 0.3 ms. This is more than an *order of magnitude* improvement over response times of conventional a-Si (~5 ms) and best reported $MoS_2$ devices (~50 ms). The van-der-waals heterojunction presented here yields a high photoresponsivity of 210 mA/W at green light-the wavelength used in commercial imaging systems. This responsivity is 4X larger than that of the best $MoS_2$ devices, and 2X larger than that of commercial a-Si devices. The 10X improvement in speed with high photoresponsivity provides a potential solution to a decades-long problem for thin film imagers and could find applications in large area electronics such as biomedical imaging and x-ray fluoroscopy.

Molybdenum disulfide ($MoS_2$), a member of the family of layered transition metal dichalcogenides, has long been used as lubricants [1]. The interest in electronic and optoelectronic applications of $MoS_2$ thin films has been kindled by recent demonstrations of high mobility transistors with $MoS_2$ active layer [2,3]. While mono-layer $MoS_2$ is targeted for high end applications such as beyond the silicon transistors [2-8], thin film $MoS_2$ could be suitable for other applications such as displays and thin film sensors [9]. Recently, thin film $MoS_2$ transistors with mobilities of 70-100 $cm^2/Vs$ were demonstrated [10,11], which could be used for driving pixels on displays. Phototransistors with thin film $MoS_2$ also showed promise for thin film photodetectors, even though their responsivity at the wavelength of 550 nm was about 50 mA/W [10]. On the other hand, among the commercially available thin film photodetectors, amorphous silicon (a-Si) is widely used as the sensing element for several applications including indirect x-ray imagers for radiology [12-15]. However, persistent photoconductivity, due to structural defects in amorphous silicon, poses several problems such as slow operation speed (tens of frames per second) and image retention (lag), making it challenging to address high speed applications such as fluoroscopy and tomography [12-15]. Therefore, a fast photodetector would expand the range of applications. We demonstrate that thin film $MoS_2$ and amorphous silicon form a diode that results in such a high speed photodetector.

**Results:**

We investigated a lateral metal-semiconductor-metal (MSM) device structure consisting of thin film $MoS_2$ covered with thin film a-Si. A schematic of the device structure is shown in Fig. 1a. The energy band diagram is shown in Fig. 1b, where an electron affinity of 3.8 and 4.3 and a bandgap of 1.6 and 1.3 eV for a-Si and $MoS_2$, respectively, was assumed [11,16,17]. In this structure, if photons are absorbed in the a-Si layer, photogenerated electrons are expected to diffuse to the a-Si/$MoS_2$ junction and to subsequently get swept into the $MoS_2$ side. In general, both layers may contribute to photocurrent. It is important to note, however, that the electrons that get transferred to $MoS_2$ should move with a higher velocity due to a larger mobility of $MoS_2$ as compared to a-Si. Bulk mobility of a-Si is about 10 $cm^2/Vs$ [18,19], whereas measured mobility of our $MoS_2$ thin films were 2-3X higher (see discussions later). These fast moving electrons in $MoS_2$ make the heterojunction lead to a much faster photoresponse for the a-Si/$MoS_2$ photodetector.

Next we discuss experimental results that support the above discussions. We made two MSM devices on $SiO_2$ substrate. The semiconductor layer consists of a 60 nm mechanically exfoliated $MoS_2$ flake covered by 100 nm a-Si in one device, and the 100 nm a-Si film alone for the second one. The schematic of device cross sections and top view photomicrographs of the fabricated devices are shown in Fig. 2a and b. The length and width of the fabricated devices are roughly 5 and 6 µm, respectively. The current-voltage (IV) characteristics of the devices were measured under dark and illuminated conditions. Figure 2c shows the dark IV characteristics. The dark current of the device with just a-Si active layer is below the noise level of our characterization system which is about 50 fA in average. Low dark currents are expected for a-Si devices as the resistivity of a-Si is above $10^{10}$ Ωcm [19]. The dark current of the hybrid a-Si/$MoS_2$ device is also below the noise level for voltages less than 0.25 V, and increases over the noise level beyond 0.25 V. It is about 0.3 pA at 1V. It should be noted that we also evaluated MSM devices with just a thin film $MoS_2$ layer, where we found that their drawback is the high dark current of about 40 pA at 1V, see Supplementary Materials and Figure S2. The measured photo IV characteristics of the hybrid a-Si/$MoS_2$ and a-Si devices are also shown in Fig. 2c. For this measurement, a fiber-coupled broadband light out of a halogen lamp was used. The optical power density was 10 mW/cm$^2$. As seen, the photocurrent of the hybrid a-Si/$MoS_2$ device is about one order of magnitude larger than that of the a-Si device.

To obtain further insight into the device operation, we measured photoresponse of devices at three wavelengths corresponding to blue, green, and red colors. For these measurements, we used standard LEDs as light sources and the incident power was 0.4 mW/cm$^2$. Photoresponsivity, calculated as the ratio of photocurrent to incident power, is shown in Fig. 3a. As seen, the responsivity of the hybrid a-Si/$MoS_2$ device is larger than that of the a-Si device for all wavelengths. For example, at $\lambda$=400 nm, they are about 150 and 60 mA/W for the hybrid a-Si/$MoS_2$ and the a-Si device, respectively. For $\lambda$=550 nm, the responsivity is maximum about 210 mA/W. This responsivity is 4X larger than that of best $MoS_2$ devices [10], and 2X larger than that of a-Si devices reported here. The green wavelength is of special interest as x-ray imagers operate at this frequency. At the wavelength of 630 nm, the photoresponse of the hybrid device is seven times larger than that of the a-Si device. The reason for this larger gain is that the a-Si layer does only absorb a part of the incident light at 630 nm and thus the underlying $MoS_2$ is also contributing to photogeneration. This is confirmed by transmittance measurements as a function of wavelength as shown in Fig. 4.

Thus far, based on the photoresponsivity data, one may conclude that the junction between the $MoS_2$ and a-Si enhances the performance in the case photogenerated electrons inside the a-Si that survive and are able to diffuse to the junction are swept to the $MoS_2$ side. To test what happens for a thicker a-Si, we made another set of devices with 300 nm a-Si with the same dimensions as those of the first set (Note that the diffusion length of a-Si is about 300 nm [19]). Fig 3b shows the photoresponsivity data measured under similar conditions like before. As seen, devices with 300 nm a-Si, with and without $MoS_2$, give the same photoresponse for all three incident lights. Therefore, in this case, photogenerated electrons completely reside inside the a-Si layer. Compared to the devices with 100 nm a-Si, the ones with 300 nm a-Si give larger photocurrents and, for example, at $\lambda$=550 nm the responsivity is about 1 A/W.

Next we measured the transient response of both set of devices by pulsing the incident light by biasing the LEDs with a 3V, 100 Hz square voltage out of a function generator. The results are shown in Fig. 3c and d. As seen from Fig. 3c, the a-Si/$MoS_2$ device with 100 nm a-Si is the fastest with rise and fall times of about 0.2-0.5 ms. For comparison, the response time of conventional a-Si and best reported $MoS_2$ devices is about 5 ms and 50 ms, respectively [14,8]. On the other hand, devices with 300 nm a-Si show similar dynamic responses with rise and fall times of about 3 ms. This observation again indicates that $MoS_2$ is not functioning once the a-Si is 300 nm thick. The signal from the device with just 100 nm a-Si, without $MoS_2$, is not regular and sometimes does not reach the final values, see Fig. 3c. Nevertheless we may estimate a response time of several ms for this device too, considering data from the three wavelengths as shown in Supplementary Fig. S5. Thus the $MoS_2$ junction is beneficial and results in the fast photoresponse when the thickness of a-Si is less than the diffusion length of electrons, so that electrons survive to reach the junction and be transferred to the $MoS_2$. Similar transient responses were measured for all three wavelengths; see Supplementary Figures S3-7. We also investigated the photoresponse of the a-Si/$MoS_2$ device, with 100 nm a-Si, when the incident power changes. Fig 3e shows the photocurrent as a function of incident power from 50 to 400 $\mu W/cm^2$. As seen, it is fairly linear versus the input power. Summarizing these results, we conclude that there exists a trade-off between speed and amplitude of photocurrent in this heterojunction. While the larger mobility of $MoS_2$ helps increase the speed of the device, a smaller thickness of a-Si, which is needed to ensure that electrons can reach the junction before recombining, could lead to reduced absorption and therefore smaller photocurrent. By varying the thickness of a-Si and $MoS_2$ layers, it should be possible to obtain an optimized scenario where a substantial increase in speed can be gained without giving up the photocurrent significantly.

Before concluding, we provide additional experimental measurements that elucidate the physics of charge transfer from a-Si into the underlying $MoS_2$ at the junction. Fig. 4a shows two sets of devices with 100 and 300 nm a-Si, with and without $MoS_2$, that were subjected to an incident light of $\lambda$=400 nm. The reason to choose this wavelength is that it is completely absorbed in the top a-Si layer without reaching the $MoS_2$. Fig. 4b shows transmission measurements of a-Si in the UV-visible range where below 420 nm the transmission is virtually zero. This implies that in the a-Si/$MoS_2$ devices the $MoS_2$ is not receiving any blue light. However, $MoS_2$ is electrically active as deduced from the IV data shown in Fig. 4c, for the devices with 100 nm a-Si. As seen, the IV of the a-Si device is linear, but that of the hybrid a-Si/$MoS_2$ device is non-linear and larger. Therefore, this validates the assumption made in the beginning that $MoS_2$ and a-Si form a diode where photogenerated electrons diffuse from a-Si towards the junction and are swept into the $MoS_2$. When the a-Si is 300 nm thick, both IVs are the same and fairly linear, see Fig. 4d. This implies that there is no charge transfer from 300 nm a-Si into the $MoS_2$ layer. In addition, we mentioned that transferred electrons move with a higher mobility in $MoS_2$ and this contributes to the larger photocurrent of the hybrid device. To support this argument, we made a bottom-gate transistor out of thin film $MoS_2$ on thermally grown $SiO_2$ as the gate dielectric. The schematic of cross section and the top view photomicrograph of the transistor are shown in Fig.

4e. The transfer characteristic is shown in Fig. 4f. A mobility of 28.6 $cm^2/Vs$ was extracted from the linear plot of the curve, as shown in the inset of Fig. 4f. This value is larger than that of the bulk of a-Si, about 10 $cm^2/Vs$ [18,19]. Thus the higher mobility of $MoS_2$ is effective in enhancing the photoresponse of the hybrid photodetector.

**Discussion:**

The a-Si/$MoS_2$ photodetector could be used for biomedical imaging applications where a fast photoresponse is required. For example, currently, flat-panel x-ray imagers based on a-Si p-i-n photodetectors operate at frame rates in the range of 10-100 Hz, limited by the slow photoresponse of a-Si. The detector presented here is much faster and can offer a speed of operation up to several kHz, considering its rise and fall times of 0.3 ms. Faster response also allows shorter x-ray exposure times to patients which helps to reduce the health hazards of x-ray radiation. From technology point of view, the advantage of this detector is its simplicity of fabrication which can be easily integrated with other components and readout circuits.

In conclusion, we demonstrated that molybdenum disulfide forms a heterojunction diode with amorphous silicon which leads to a significantly fast dynamic photoresponse. The response time of the hybrid a-Si/$MoS_2$ photodetector is over 10X and 100X faster than that of the commercially available amorphous silicon and best reported $MoS_2$ devices, respectively. The device has a high photoresponsivity of about 210 mA/W at the wavelength of 550 nm. In addition, it has a low dark current of 0.3 pA at 1V bias. Going forward, such devices could be integrated with $MoS_2$ transistors, as readout circuits and pixel amplifiers, leading to a monolithic large area technology based on layered semiconductors.

**Methods:**

Flakes of $MoS_2$ were exfoliated on $SiO_2$ substrates from a piece of $MoS_2$ crystal (provided by the SPI Supplies, www.2spi.com) using the scotch-tape mechanical cleavage method. Subsequently, the thin film $MoS_2$ flakes were coated with amorphous silicon (a-Si) thin films, deposited by plasma-enhanced chemical vapour deposition (PECVD) using silane gas source at 260 C. Metal contacts were formed by the conventional lift-off technique, where Ti (20 nm) and Au (160 nm) were deposited by electron-beam evaporation at room temperature. Current-voltage characteristics of devices were measured by an Agilent 4155C semiconductor parameter analyzer, under dark and illuminated conditions. For the latter, a fiber-coupled broadband light out of a halogen lamp was used. The optical power density was about 10 $mW/cm^2$. We also measured photoresponse of devices at three wavelengths corresponding to blue, green, and red colors. For these measurements, we used standard LEDs as light sources and the incident power was about 0.4 $mW/cm^2$.

## Author Contributions

M.E. and S.S. devised the device concept and initiated the research. M.E. worked on device fabrication, performed measurements, and wrote the manuscript. All the authors read and commented on the manuscript.

## Additional Information

**Competing financial interests**: the authors declare no competing financial interests.

**Figure Captions:**

Figure 1: Operation concept of molybdenum disulfide amorphous silicon heterojunction photodetector. (**a**) schematic and (**b**) energy band diagram of a-Si/$MoS_2$ heterojunction MSM photodetector. The light is incident from a-Si side, optical absorption occurs in a-Si, and photogenerated electrons diffuse to the underlying $MoS_2$ layer and are transferred across the $MoS_2$ layer toward a metal contact.

Figure 2: (**a**) Schematic of cross sections of two metal-semiconductor-metal (MSM) photodetectors and (**b**) corresponding top view photomicrograph of devices. The length and width of devices is about 5 and 6 µm, respectively. (**c**) Measured dark and photo current-voltage (IV) characteristics. The semiconductor layer is 100 nm a-Si film (labeled as a-Si), and 60 nm $MoS_2$ flake covered by 100 nm a-Si (labeled as a-Si/$MoS_2$), see Supplementary Materials and Figure S1 for further details.

Figure 3: Measured photoresponsivity of a-Si/$MoS_2$ and a-Si MSM devices for three wavelengths corresponding to blue, green, and red colors; (**a**) and (**b**) when the thickness of a-Si is 100 and 300 nm, respectively. Photocurrents at the applied voltage of 1V were taken for calculations. The incident power was 0.4 $mW/cm^2$. (**c**) and (**d**) transient response of the two sets of devices for the incident wavelength of 550 nm. For part (**c**), the voltage on the a-Si device was increased to obtain photocurrent pulse amplitudes comparable to that of the hybrid device. (**e**) photocurrent of the a-Si/$MoS_2$ device as a function of incident power. Applied voltage was 1V. In parts (**a**) and (**b**), lines are for the eye guide only.

Figure 4: Experimental data that support the proposed concept of operation of the photodetector. (**a**) Schematic diagram of devices subjected to blue incident light. (**b**) Measured UV-visible transmission of 100 and 300 nm a-Si thin films, used in the two sets of devices, with and without $MoS_2$ flakes. (**c**) Measured photo IV of devices with 100 nm a-Si, showing that the IV of the a-Si/$MoS_2$ device is non-linear and $MoS_2$ is effective in boosting the photocurrent. (**d**) Measured photo IV of devices with 300 nm a-Si, showing that $MoS_2$ is not effective in enhancing the photocurrent. (**e**) schematic cross section and top view photomicrograph of the bottom-gate transistor, made of a 73 nm thick $MoS_2$ flake. (**f**) Measured transfer characteristic where a mobility of 28.6 $cm^2/Vs$ was extracted.

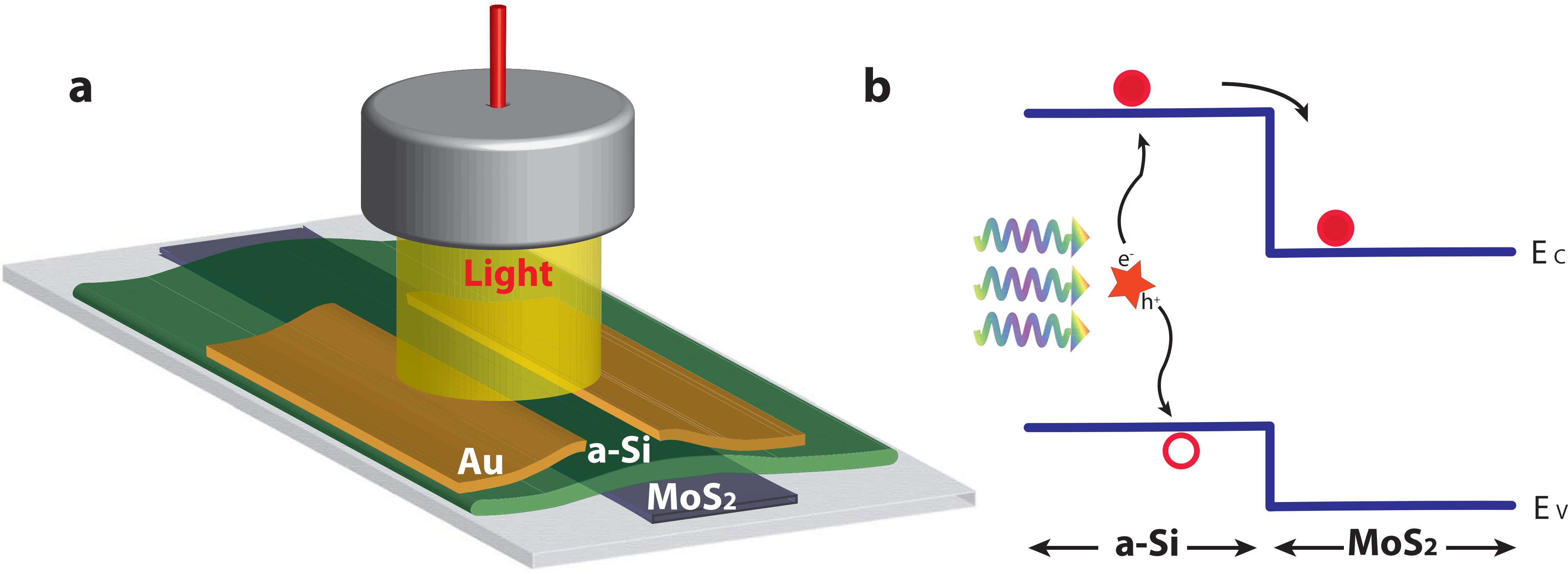

**Figure 1: operation concept of molybdenum disulfide amorphous silicon heterojunction photodetector.**

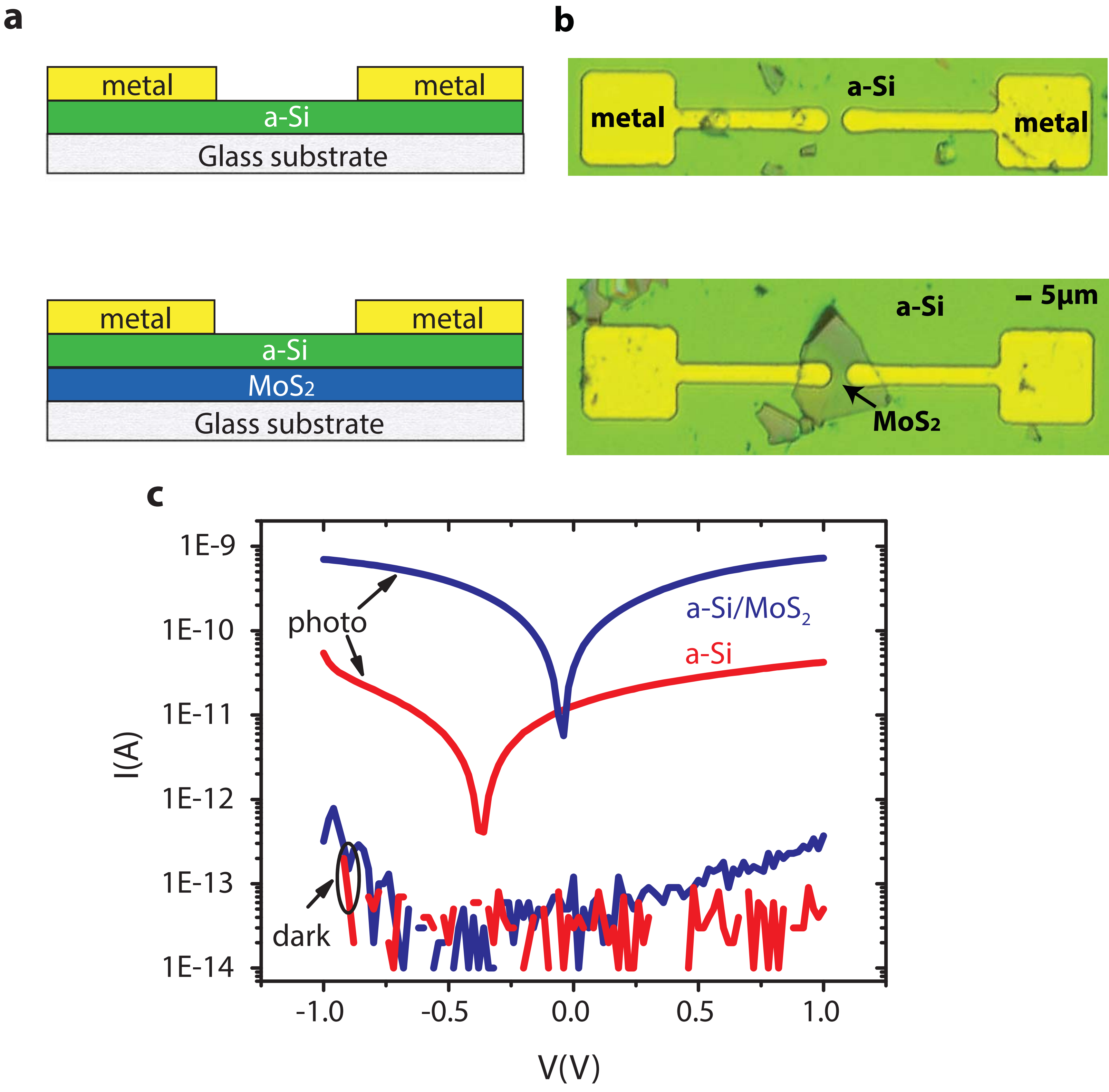

**Figure 2: schematic of cross sections and photomicrograph of fabricated devices.**

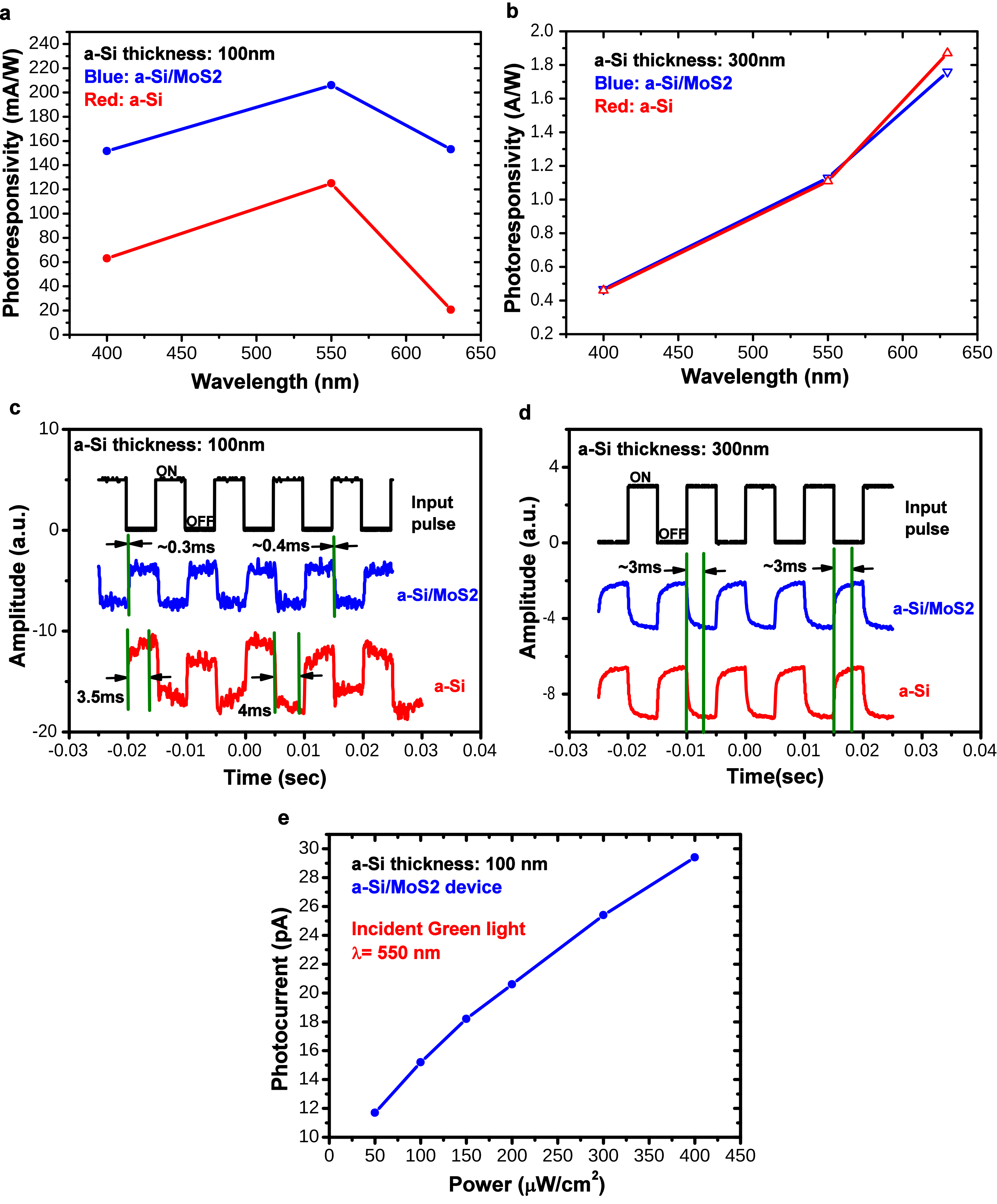

Figure 3: characterization results of MSM photodetectors.

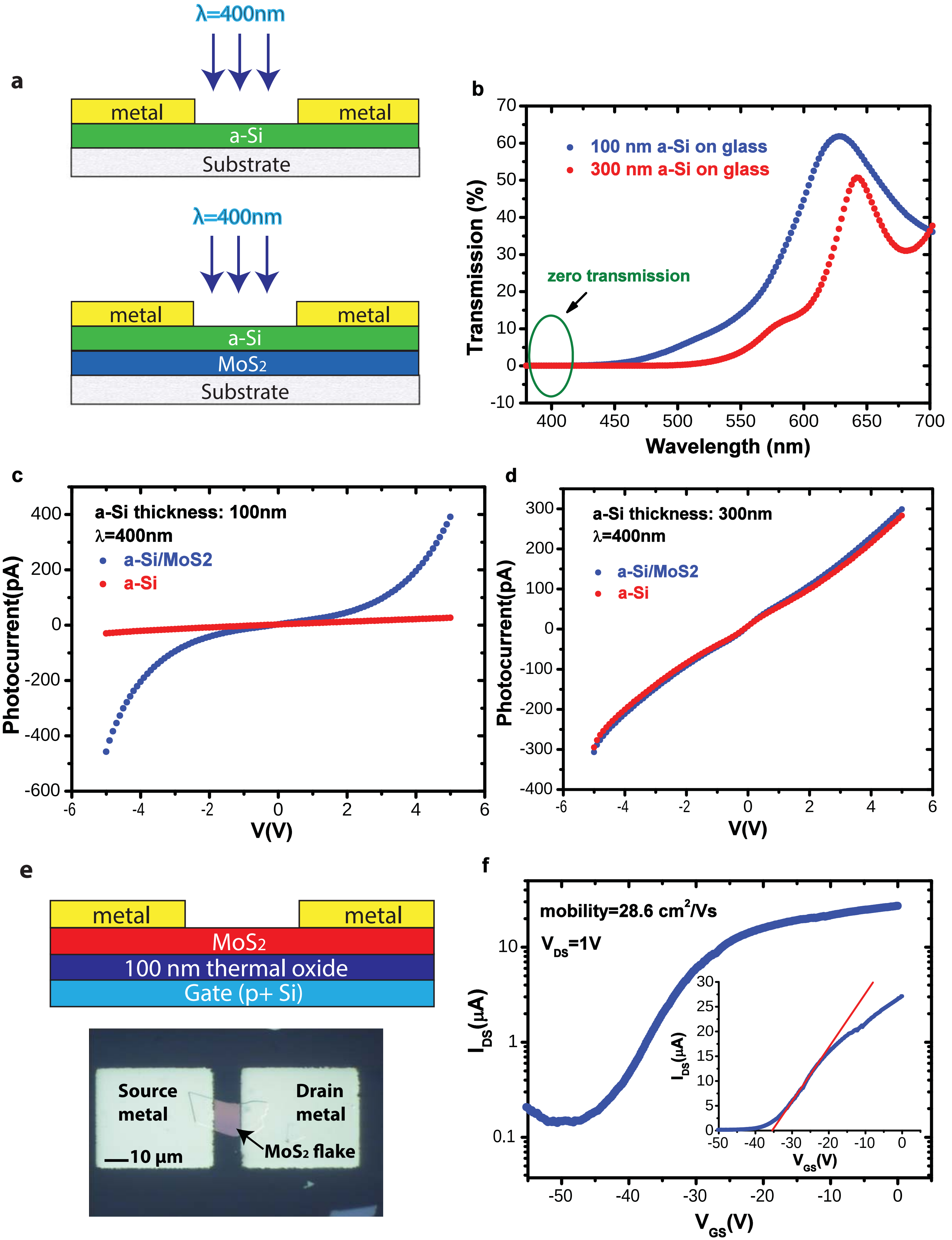

Figure 4: experimental data that support the proposed concept of operation of the photodetector.

# Supplementary Materials for

## High Performance Molybdenum Disulfide Amorphous Silicon Heterojunction Photodetector

Mohammad R. Esmaeili-Rad and Sayeef Salahuddin

Department of Electrical Engineering and Computer Sciences, University of California, Berkeley, California 94720, United States

### Device Fabrication and Measurement:

Flakes of $MoS_2$ were exfoliated on $SiO_2$ substrates from a piece of $MoS_2$ crystal (provided by the SPI Supplies, www.2spi.com) using the scotch-tape mechanical cleavage method. Subsequently, the thin film $MoS_2$ flakes were coated with amorphous silicon (a-Si) thin films, deposited by plasma-enhanced chemical vapour deposition (PECVD) using silane gas source at 260 C. Metal contacts were formed by the conventional lift-off technique, where Ti (20 nm) and Au (160 nm) were deposited by electron-beam evaporation at room temperature. Current-voltage characteristics of devices were measured by an Agilent 4155C semiconductor parameter analyzer, under dark and illuminated conditions. For the latter, a fiber-coupled broadband light out of a halogen lamp was used. The optical power density was about 10 mW/cm$^2$. We also measured photoresponse of devices at three wavelengths corresponding to blue, green, and red colors. For these measurements, we used standard LEDs as light sources and the incident power was about 0.4 mW/cm$^2$. The thickness of the $MoS_2$ flakes was measured using a mechanical, stylus-based step profiler, Alpha-Step IQ Surface Profiler. Figure S1 shows the top view photomicrograph of the hybrid a-Si/$MoS_2$ metal-semiconductor-metal (MSM) photodetector and a lateral line scan of the $MoS_2$ flake, showing that the thickness of flake is about 60 nm.

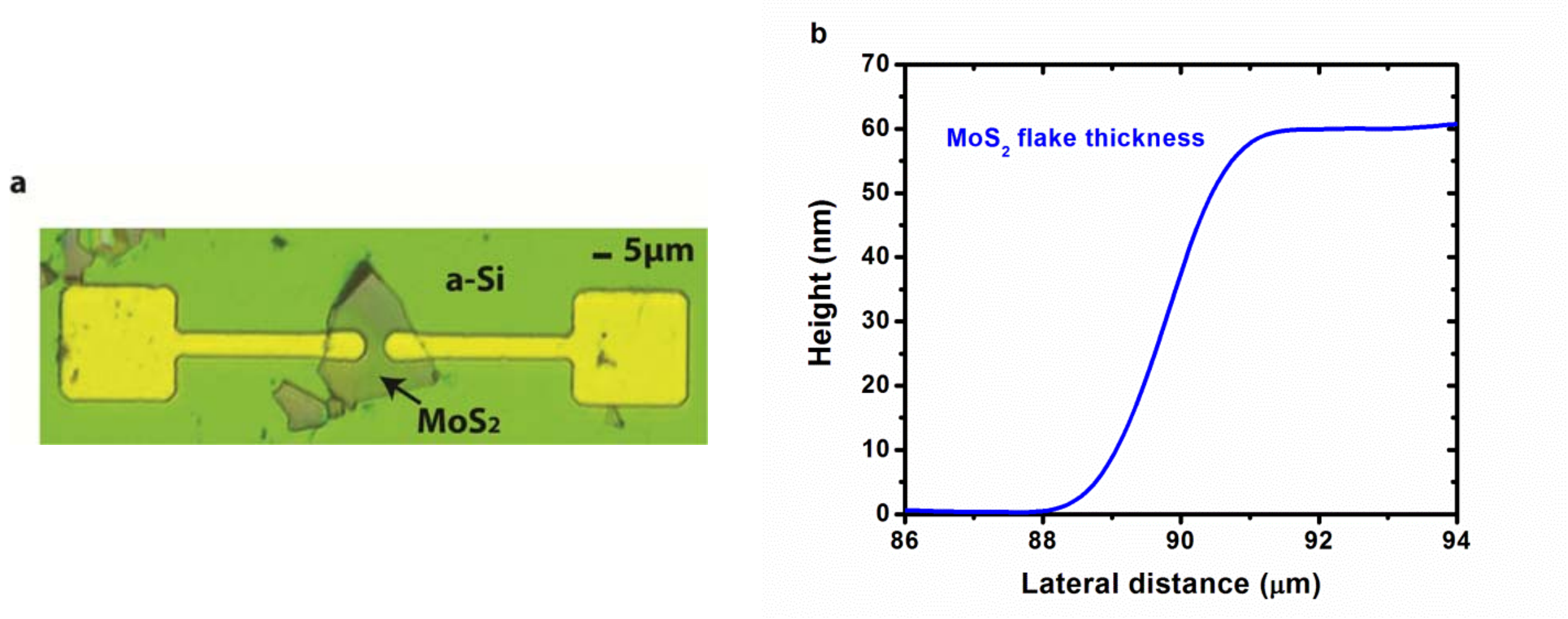

Figure S1: **a**, top view photomicrograph of the a-Si/$MoS_2$ MSM photodetector. **b**, lateral line scan of the $MoS_2$ flake, showing that the thickness of flake is about 60 nm.

**High Dark Current of metal-semiconductor-metal (MSM) photodetectors with just thin film $MoS_2$:**

In the beginning of this research, we studied thin film $MoS_2$ as an active layer for MSM photodetectors. Figure S2a and b show the schematic of the device cross section and the top view photomicrograph of the fabricated device, respectively. The device length and width is about 5 and 6 µm, respectively. The current-voltage (IV) characteristics of the device were measured under dark and illuminated conditions. For the latter, a fiber-coupled broadband light out of a halogen lamp was used. The optical power density was about 10 mW/cm$^2$. Figure S2c shows the IV characteristics.

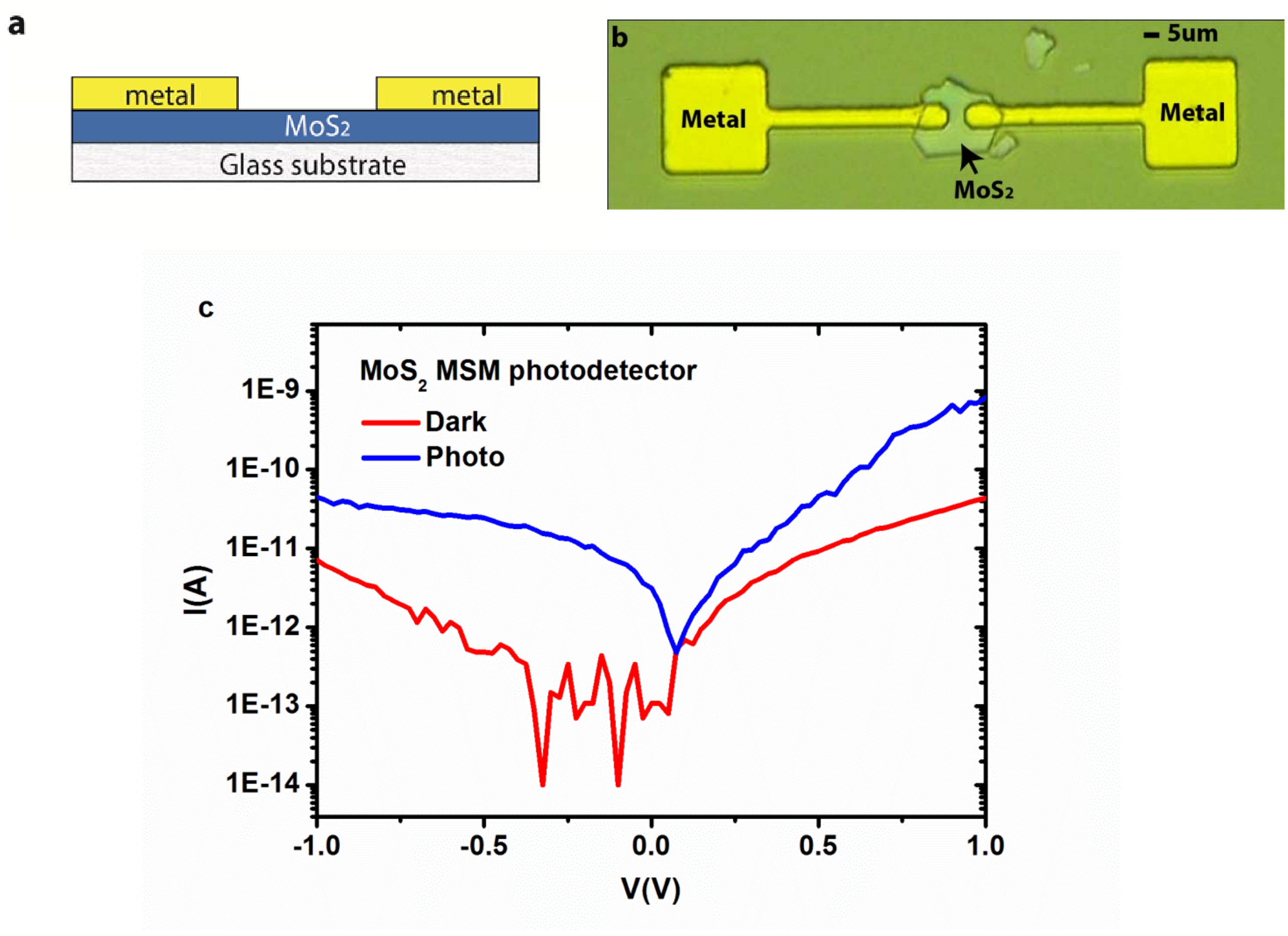

Figure S2: **a**, schematic of cross section of the $MoS_2$ MSM device, and **b**, the top view photomicrograph of the fabricated device. **c**, dark and photo current-voltage (IV) characteristics. For the latter, a fiber-coupled broadband light out of a halogen lamp was used. The optical power density was about 10 $mW/cm^2$.

As seen, the drawback of the MSM device, with just $MoS_2$ active layer, is the high dark current. For example, at 1V applied voltage, the dark current is about 40 pA. This value is about two orders larger than the dark current of the a-Si/$MoS_2$ device presented in the main text. The high dark current results in a low dynamic range which limits the application of $MoS_2$ MSM photodetector in practical systems. Therefore, we found that a-Si layer helps to reduce the dark current, see the discussions in the main text. From Fig. S2c, one may also notice that both photo and dark IV curves are asymmetric with respect to the applied voltage, as the currents for identical positive and negative biases are not the same. For example, the photo currents are about 1nA and 40pA at +1 and -1V, respectively. This could be due to some variations or asymmetry in the Schottky barriers between the metal contacts and the $MoS_2$ layer.

**Measurement Setup for Transient Photoresponse:**

Figure S3 shows the schematic of the setup used for measurements of the transient photoresponse of devices. The LED light source was biased by a 100 Hz square voltage out of a function generator. The light pulse was incident on the device under test (DUT) from top side. A diffuser was used to make sure the light intensity is uniform across a wide area of several $cm^2$, thus making sure the DUT is receiving uniform light intensity. The distance between the LED and the device was about 4 cm to obtain uniform illumination. The DUT was biased by a DC voltage to extract photo current pulses. The photocurrent was amplified by a current to voltage (I to V) amplifier to obtain a voltage output. Finally, the voltage signal was fed to an oscilloscope along with the reference (sync) signal out of the function generator.

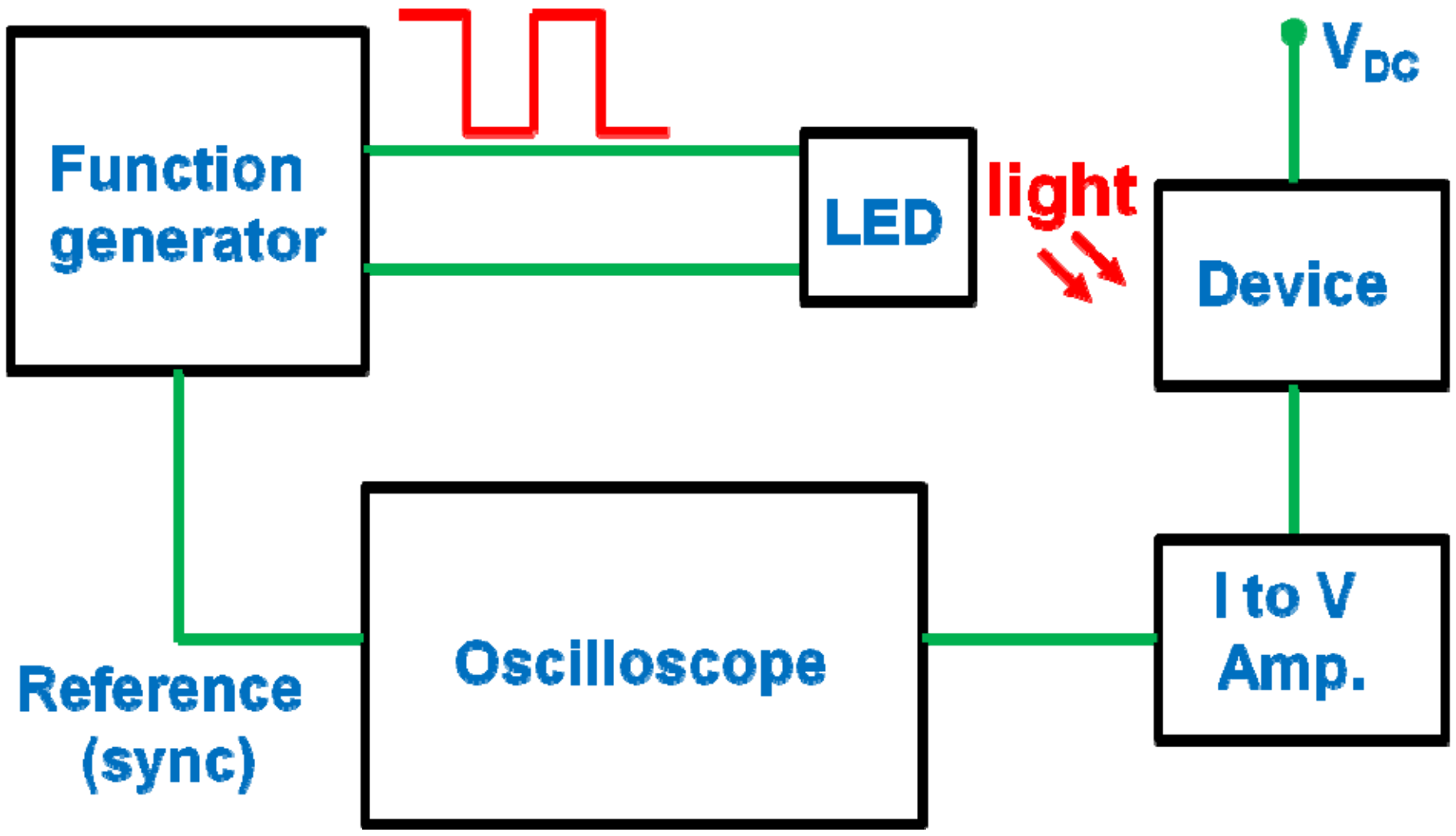

Figure S3: schematic of the setup used for measurements of the transient photoresponse of devices.

Using the above setup, we measured the transient photoresponse of the two sets of devices, four different devices in total. Those are devices with 100 and 300 nm a-Si, with and without $MoS_2$. The applied DC voltage was 1V for most of measurements, except for devices with just 100 nm a-Si, where we had to increase the applied voltage to 5-10 V to obtain a less noisy signal and to make the output signal amplitude comparable to that of the hybrid a-Si/$MoS_2$ device. The results are summarized in Figures S4-7. As seen from Fig. S4, the a-Si/$MoS_2$ hybrid device gives the fastest response for all three incident lights. We estimated rise and fall times of about 0.2-0.5 ms for this device. The signal generated by the a-Si device, 100 nm thick, is not regular and sometimes does not reach to a stable level, see Fig. S5. We have observed this behavior for several devices. Nevertheless, we can estimate rise and fall times of 3-4 ms based on the data presented in Fig. S5. The slow response of a-Si devices is attributed to charge trapping and detrapping dynamics which happens in the disordered material structure of amorphous silicon.

For the two devices with 300 nm amorphous silicon, with and without $MoS_2$, one can see that the transient responses are virtually identical, see Figs. S6 and S7. We estimated rise and fall times of about 2-3 msec for these two devices, irrespective of the presence of the $MoS_2$ flake. Therefore, $MoS_2$ is not functioning in the hybrid device with 300 nm a-Si. In other words, once the amorphous silicon is thicker than the diffusion length of electrons in amorphous silicon, then electrons tend to move laterally inside the top a-Si layer without crossing the junction and being transferred to the $MoS_2$ layer.

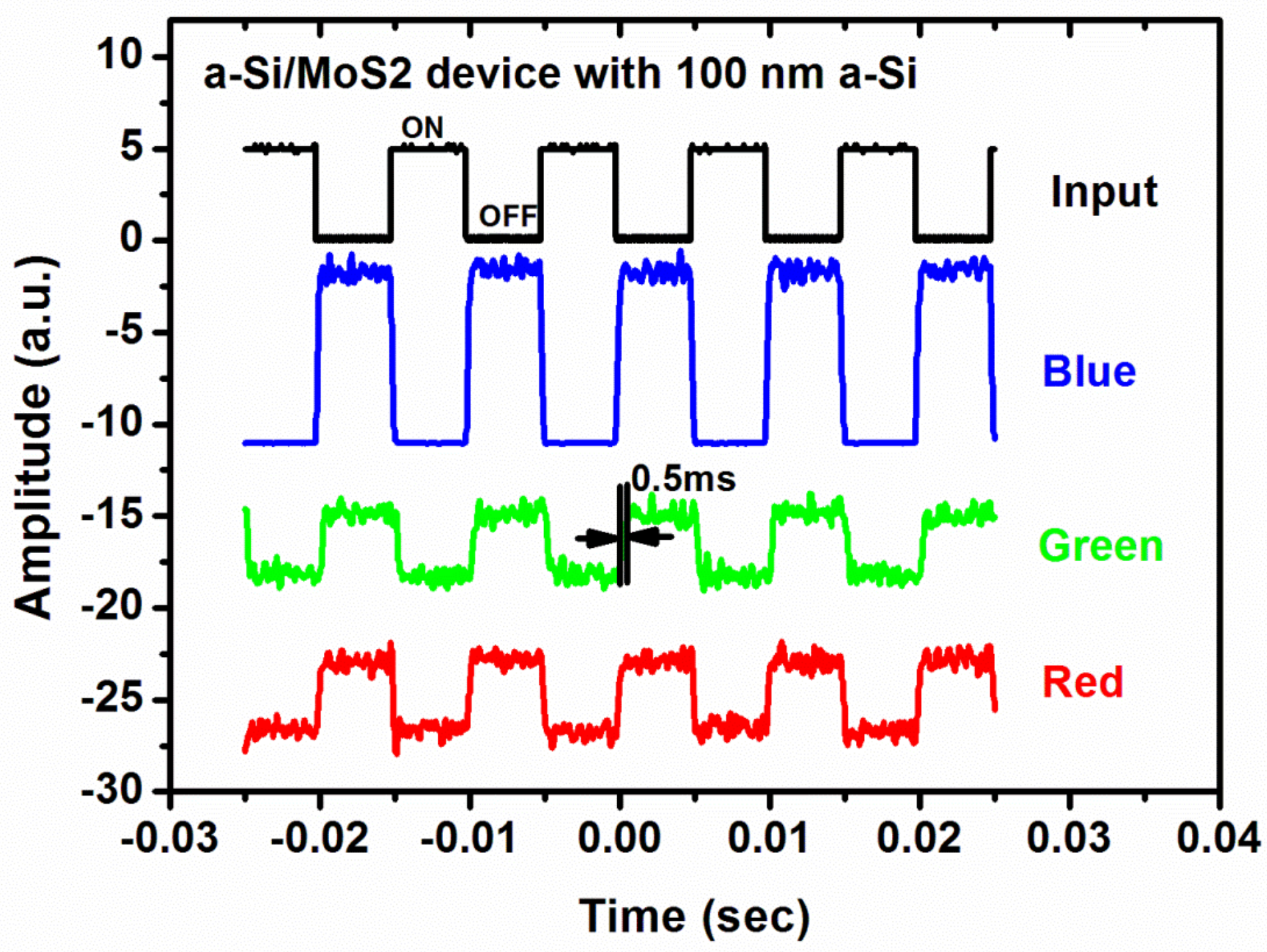

Figure S4: Transient photoresponse of the a-Si/$MoS_2$ MSM device with 100 nm a-Si for three wavelengths corresponding to blue, green, and red wavelengths. The curves are shifted vertically for display purposes.

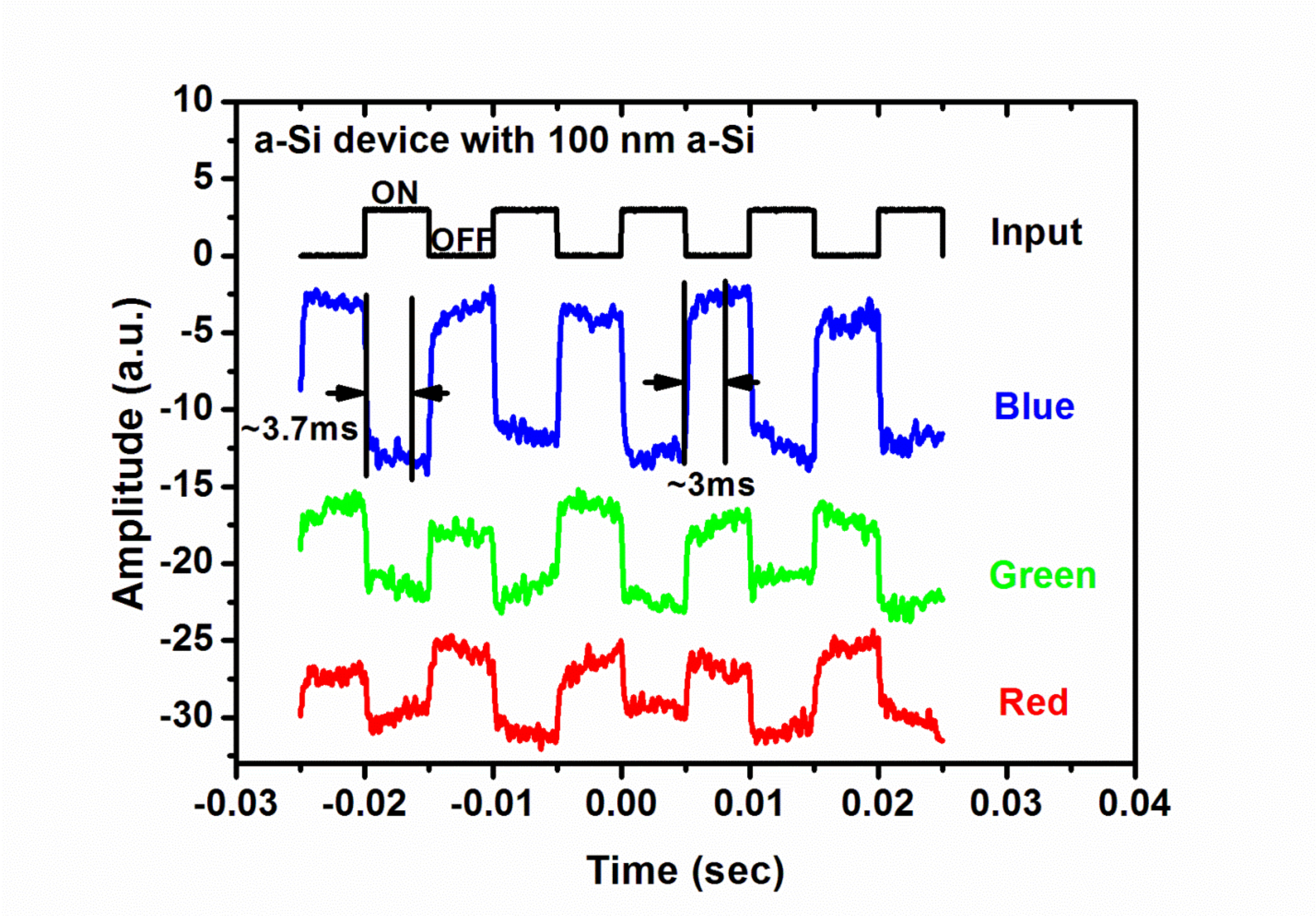

Figure S5: Transient photoresponse of the a-Si MSM device with 100 nm a-Si for three wavelengths corresponding to blue, green, and red wavelengths. The curves are shifted vertically for display purposes.

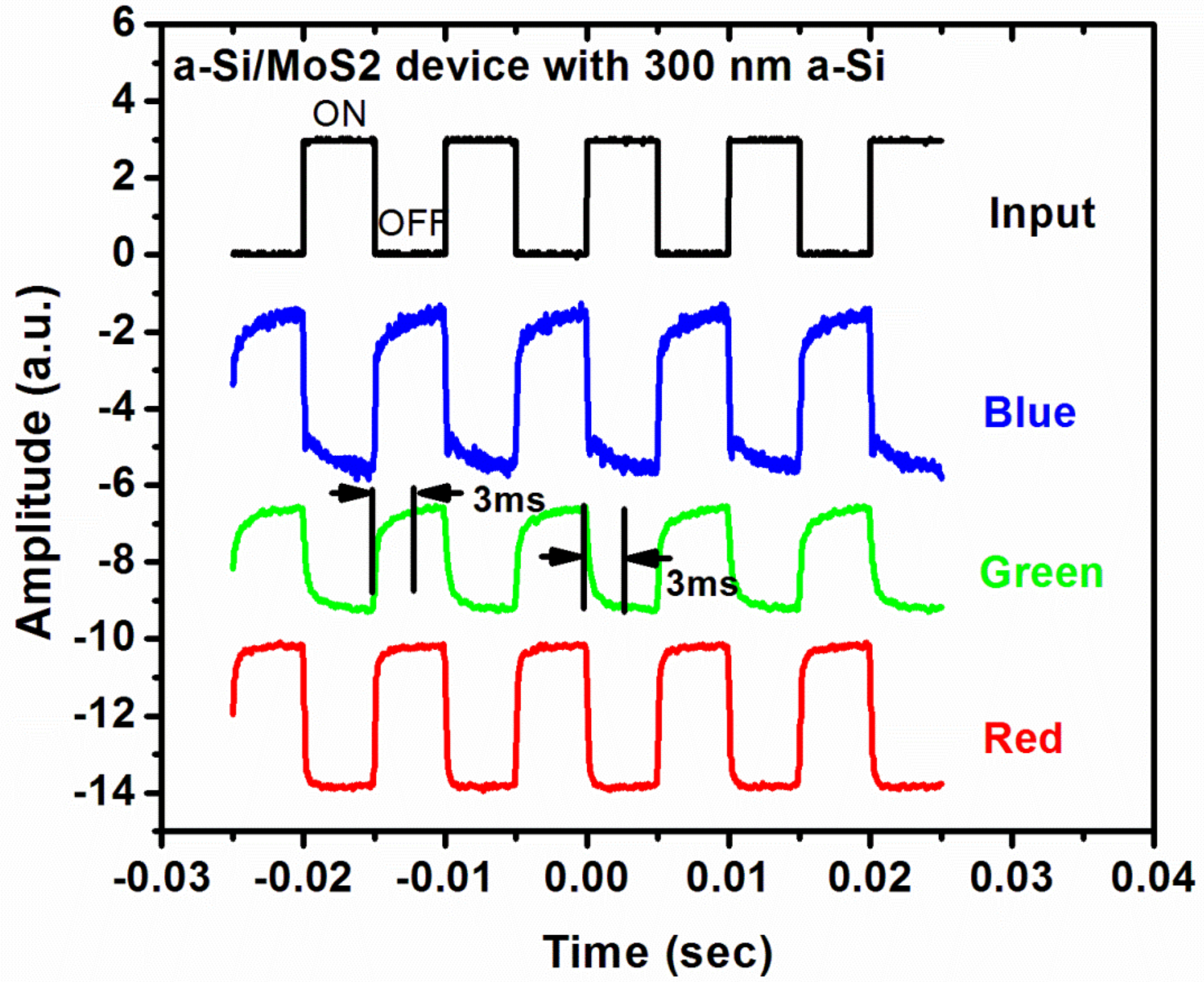

Figure S6: Transient photoresponse of the a-Si/$MoS_2$ device with 300 nm a-Si for three wavelengths corresponding to blue, green, and red wavelengths. The curves are shifted vertically for display purposes.

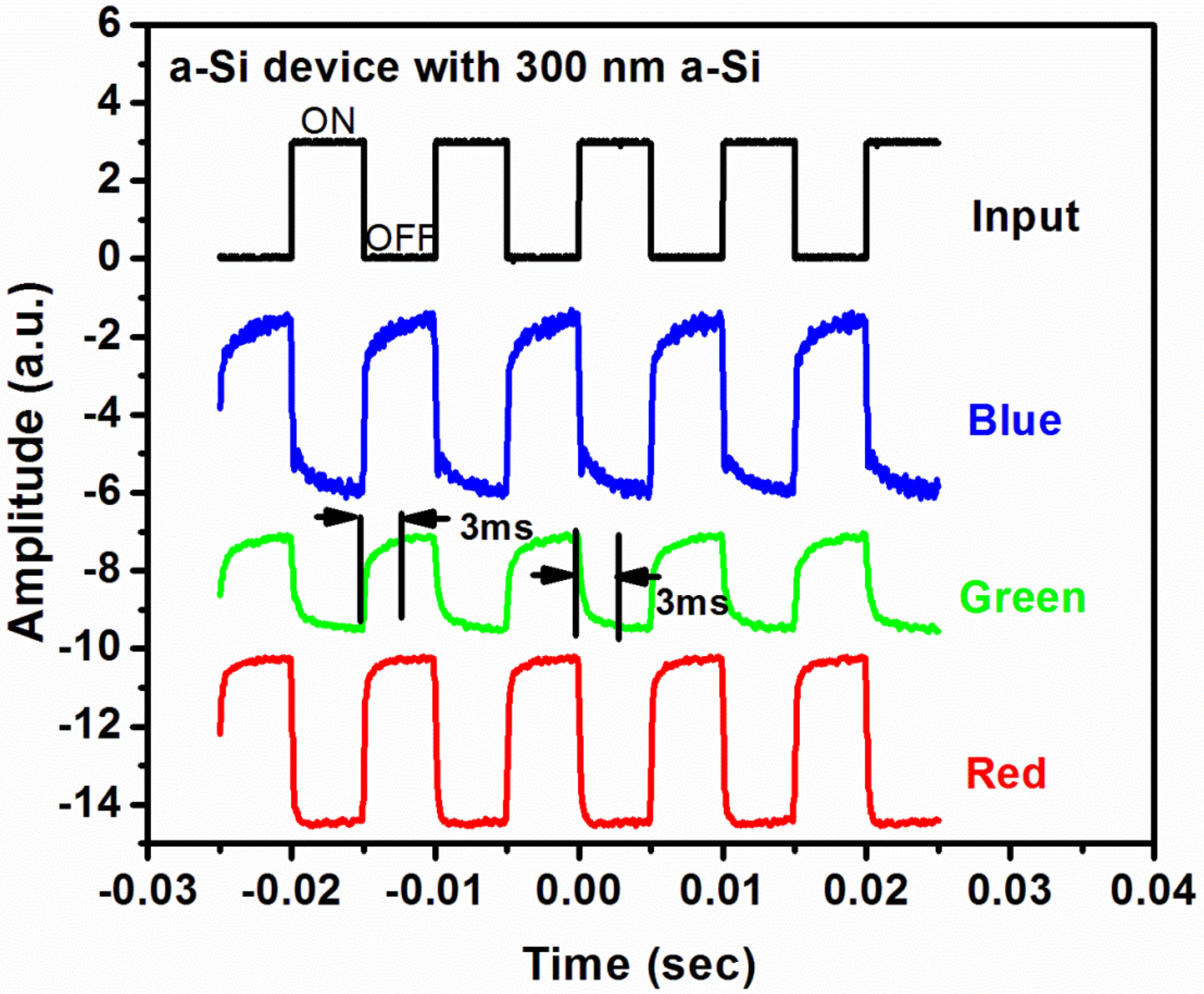

Figure S7: Transient photoresponse of the a-Si MSM device with 300 nm a-Si for three wavelengths corresponding to blue, green, and red wavelengths. The curves are shifted vertically for display purposes.